\documentclass[journal,twoside]{IEEEtran}
\usepackage{cite}
\usepackage{array}
\usepackage{graphicx}
\usepackage{amsmath,amssymb,amsfonts,bm}
\usepackage{algorithm,algorithmic}
\usepackage{multirow,booktabs,threeparttable}
\usepackage{textcomp}
\usepackage{xcolor}

\ifCLASSOPTIONcompsoc
\usepackage[caption=false,font=normalsize,labelfont=sf,textfont=sf]{subfig}
\else
\usepackage[caption=false,font=footnotesize]{subfig}
\fi

\begin{document}
	
\title{Range-Doppler Sidelobe Suppression for Pulsed Radar Based on Golay Complementary Codes}

\author{Zhong-Jie~Wu, Chen-Xu~Wang, Pei-He Jiang, and~Zhi-Quan~Zhou%
	\thanks{Manuscript received 15-Jan-2020; revised 00-00-0000; accepted 00-00-0000. Date
	of publication 00-00-0000; date of current version 00-00-0000. The associate editor coordinating the review of this manuscript and approving it for publication was Dr. XX XX. This work was supported by National Natural Science Foundation of China under Grant 61771160. {\textit{(Corresponding author: Zhi-Quan Zhou.)}}}
	\thanks{Z. Wu, C. Wang, and Z. Zhou are with School of Information Science and Engineering, Harbin Institute of Technology, Weihai 264209, China (e-mail: hit\_wzj@163.com, wangchenxu@hit.edu.cn, zzq@hitwh.edu.cn).}
	\thanks{P. Jiang is with School of Opto-Electronic Information Science and Technology, Yantai University, Yantai 264005, China (e-mail: jiangpeihe@ytu.edu.cn).}
	\thanks{Digital Object Identifier 0000000000}\vspace{-2em}}

\markboth{IEEE Signal Processing Letters,~Vol.~14, No.~8, August~2015}%
{Wu \MakeLowercase{\textit{et al.}}: On the Numeric Construction of Complete Complementary Codes}

\maketitle

\begin{abstract}
	To relieve the interference caused by range-Doppler sidelobes in pulsed radars, we propose a new method to construct Doppler resilient complementary waveforms based on Golay codes. We design both the transmit pulse train and the receive pulse weights, so that the similarity between the pulse weights and a given window function is maximized and the constraints on Doppler null points and energy are met. That is summarized as a two-way partitioning problem, and then solved by semidefinite programming and randomization techniques. The novel waveform thus obtained has its range sidelobe outright suppressed in multiple and flexibly-adjustable  Doppler zones, and performs well in Doppler sidelobe suppression, Doppler resolution and SNR. It shows great promise in detecting slightly-moving weak targets with the existence of dense interference.    
\end{abstract}

\section{Introduction}
\IEEEPARstart{A}{periodic} autocorrelation functions (ACF) reveal numerous inherent properties of radar waveforms, e.g., range resolution and anti-interference capability \cite{Levanon2004}. Ideally, an ACF should be a Kronecker delta function, which is prohibited from ever existing by mathematical principles. The best one can expect is to suppress the sidelobes in particular range zones\cite{Sira2007,Cui2017,Zhao2017,Biskin2019}. Such waveforms only show their strengths in so-called cognitive radars, which feature receiver-to-transmitter feedback and transmitter adjustability\cite{Greco2018, Gurbuz2019}. This radar starts off by transmitting a traditional waveform to perceive reflectors of no interest and identify putative targets. According to their relative positions, a special waveform is designed to prevent the strong reflector sidelobes from masking the weak targets, and the final decision on the target existence is made\cite{Sira2007}. This ``enquire-verify'' mode greatly complicates the radar system, and merely brings limited benefit. 

The pulse-to-pulse waveform agility based on Golay complementary pairs (GP) can achieve outright range sidelobe~suppression and greatly simplify the above working mode. The ACFs of the sequences in a GP sum up to 0 at any out-of-phase lags \cite{Golay1949}. However, when transmitted as a coherent pulse train, a GP lost its complementary~property and possesses fairly high sidelobes in the presence of even a tiny inter-pulse Doppler shift\cite{Calderbank2009,Levanon2017}. This has spawned the concept of Doppler resilient complementary waveforms (DRCW), whose range sidelobes are suppressed in a specific Doppler zone. DRCW construction methods basically fall into two categories. The first kind repeats the constituent sequences of a GP in the order of Prouhet-Thue-Morse (PTM) codes \cite{Pezeshki2008} or first-order Reed-M\"uller (RM) codes \cite{Suvorova2007}. Meanwhile, the second kind, represented by the binomial design (BD) method\cite{Dang2011}, jointly designs the transmit pulse trains and receive pulse weights so as to improve the Doppler resilience at a cost of the loss in SNR and Doppler resolution. In pursuit of the tradeoff between Doppler resilience and Doppler resolution, \cite{Zhu2017,Zhu2018,Zhu2019} combine the outputs of the RM method and the BD method with point-wise processing. However, existing methods fail to provide adequate Doppler resilience as practical applications sometimes expect. Besides, none of them give comprehensive consideration of various metrics including Doppler sidelobe suppression, Doppler resolution and SNR.              

In this letter, we propose a novel transmit-receive design method of DRCWs. We first formulate the design problem by maximizing the similarity between the pulse weight vector and a given window function under Doppler null point and energy constraints, then approximately solve it with semidefinite programming followed by a randomization procedure, and finally verify the effectiveness of the method through numerical experiments. Our method can design a DRCW with multiple range sidelobe blanking areas on the range-Doppler plane, suppressed Doppler sidelobes, and acceptable loss in SNR and Doppler resolution. 

\emph{Notations:} Italic, bold lowercase and uppercase letters stand
scalars, column vectors and matrices. ${(\cdot)^*}$, ${(\cdot)^T}$,$\left\| {\cdot} \right\|$, ${\rm{abs}}(\cdot)$, ${\rm{sign}}(\cdot)$, ${\rm {Diag}(\cdot)}$ and ${\mathrm {tr}(\cdot)}$  denote complex conjugate, transpose, $l_2$-norm, absolute value, signum function, diagonal matrix constructed from a vector and matrix trace, respectively.

\section{Signal Model}
The binary sequences ${{\bf{x}}_i} =\left[{{x_i}[0], \cdots, {x_i}[N-1] }\right]^T, i = 1,2$ are termed a Golay pair if
\begin{equation}
	{R_{{\bf{x}}_1}}\left[ k \right] + {R_{{\bf{x}}_2}}\left[ k \right] = 2N\delta \left[ k \right],
\end{equation} 
where ${R_{{\bf{x}}_i}}\left[ k \right]$ is the discrete ACF of ${\bf{x}}_i$ at lag $k$, and $\delta \left[ k \right]$ is the Kronecker delta function. An analogue signal $x_i\left(t\right)$ is generated by using ${\bf x}_i$ to modulate a train of sub-pulses 
\begin{equation}
	x_i\left( t \right) = \sum\limits_{n = 0}^{N - 1} {{x_i}}\left[n\right] \Omega \left( {t - n{T_c}} \right),
\end{equation}  
where $T_c$ is the chip width and $\Omega\left(t\right)$ is the sub-pulse of unit energy and support set $\left[0,T_c\right]$.

Let the binary vector ${\bf s}\!=\!\left[s_0,\!\cdots,\!s_{M-1}\right]^T$ be the order of the transmitted pulses, and ${\left\{ {1, - 1,1, - 1,\cdots} \right\}}$ be the standard order. The length-$M$ coherent pulse train $\left\{{\bf x}^{\left(0\right)},\cdots,{\bf x}^{\left(M-1\right)} \right\}$ is determined by $\bf s$: if $s_m=1$, ${\bf x}^{\left(m\right)}={\bf x}_1$, otherwise ${\bf x}^{\left(m\right)}={\bf x}_2$. Hence, the corresponding baseband signal is
\begin{equation}
	u_{\mathcal{S}}\left( t \right) = \sum\limits_{m = 0}^{M - 1} {\frac{{1\! +\! {s_m}}}{2}x_1\left( {t \!-\! mT} \right) + } \frac{{1\!-\!{s_m}}}{2}x_2\left( {t\! -\! mT} \right)
\end{equation}
where $T$ is the pulse repetition interval (PRI). Let the positive-valued vector ${\bf w}=\left[w_0,\cdots,w_{M-1}\right]^T$ be the weight vector of received pulses. The weighted version of $u_{\mathcal S}\left(t\right)$ is given by
\begin{equation}
	u_{\mathcal{W}}\left( t \right)\! = \! \sum\limits_{m = 0}^{M - 1} w_m\left\{ {\frac{{1\! +\! {s_m}}}{2}x_1\left( {t \!-\! mT} \right) \!+\! \frac{{1\!-\!{s_m}}}{2}x_2\left( {t\! -\! mT} \right)} \right\}.
\end{equation}
The cross-ambiguity function of $u_{\mathcal S}$ and $u_{\mathcal W}$,  i.e., the temporal response of the receive filter $u_{\mathcal W}^*\left(-t\right)$ fed on $u_{\mathcal S} \left(t\right){\rm e}^{j\omega t}$, is defined as follows 
\begin{equation}
	{\chi _{\mathcal{S,W}}}\left( {\tau ,\omega  } \right) = \int_{ - \infty }^{ + \infty } {{u_{\mathcal S}}\left( t \right)u_{\mathcal{W}}^*\left( {t - \tau } \right){{\rm{e}}^{ j\omega  t}}dt}.
\end{equation}
${\chi _{\mathcal{S,W}}}\left( {\tau ,\omega  } \right)$ looks similar to a nail board, as it consists of a central lobe and a cluster of aliasing lobes. By selecting the radar parameters appropriately,\footnote{That is to say,  the maximum detectable range of the radar $d_{max}\le \frac{cT}{2}$ and the maximum Doppler frequency of the target $|\omega_{max}| \le \frac{\pi}{T} \ll \frac{2\pi}{NT_c}$. Those are common assumptions in pulse-Doppler radars.} the aliasing in both range and Doppler is avoidable, and the intra-pulse Doppler effect is negligible. The central lobe of ${\chi _{\mathcal{S,W}}}\left( {\tau ,\omega  } \right)$ primarily depends on the discrete-time composite ambiguity function (CAF)
\begin{subequations}
	\label{equ:CAF}
	\begin{align}
		{R_{\mathcal{S,W}}}\left( {k,\theta } \right) ={}&\sum\limits_{m = 0}^{M - 1} {{w_m}{R_{{{\bf{x}}^{\left( m \right)}}}}\left[ k \right]} {{\rm{e}}^{j\theta m}} \nonumber \\
		\label{equ:CAF1} ={}& \frac{1}{2}\left( {{R_{{\bf{x}}_1}}\left[k \right] + {R_{{\bf{x}}_2}}\left[ k \right]} \right)\sum\limits_{m = 0}^{M - 1} {{w_m}{{\rm{e}}^{j\theta m}}}  + \\
		\label{equ:CAF2} &\frac{1}{2}\left( {{R_{{\bf{x}}_1}}\left[k \right] - {R_{{\bf{x}}_2}}\left[ k \right]} \right)\sum\limits_{m = 0}^{M - 1} {{s_m}{w_m}{{\rm{e}}^{j\theta m}}}
	\end{align}
\end{subequations}   
where $\theta=\omega T$ is the Doppler shift across a PRI. Since ${\bf{x}}_1$ and ${\bf{x}}_2$ are complementary, (\ref{equ:CAF1}) vanishes at any nonzero $k$  and shapes the Doppler profile ${\chi _{\mathcal{S,W}}}\left( {0 ,\omega  } \right)$. Meanwhile, (\ref{equ:CAF2}) causes the range sidelobes and  shapes the range profile of ${\chi _{\mathcal{S,W}}}\left( {\tau ,\omega  } \right)$ at the Doppler shift $\omega$. Four performance metrics are adopted to evaluate the DRCW determined by $\left\{\bf {s,w}\right\}$. 

\subsubsection{Doppler resilience} This is described by the range~sidelobe blanking area (RSBA). We define RSBA as a Doppler shift interval within which the peak range sidelobe level (PRSL) is lower than $-60{\ \rm{dB}}$. 
\subsubsection{Doppler resolution} This is negatively correlated with the Doppler mainlobe (the mainlobe of ${R_{\mathcal{S,W}}}\left( {0,\theta } \right)$) width. The Doppler mainlobe widening ratio (DMBR) is defined as the percent increase in the -3dB mainlobe width of ${\bf w}$-weighted pulse train relative to the unweighted one.   
\subsubsection{Doppler sidelobe suppression} This is evaluated by the peak Doppler sidelobe level (PDSL).
\subsubsection{SNR} In white Gaussian noise, the coherent accumulation gain of $M$ pulses is ${{\left\| {\bf{w}} \right\|_1^2}}/{{\left\| {\bf{w}} \right\|_2^2}}$, which reaches its maximum value $M$ at ${\bf w}=\alpha{\bf 1}_M$ (i.e., the unweighted case). We define the normalized accumulation gain (NAG) as      
\begin{equation}
	NAG=10 {\rm{log}}_{10} \frac{{\left\| {\bf{w}} \right\|_1^2}}{{M\left\| {\bf{w}} \right\|_2^2}}
\end{equation}

\section{Algorithm Development}
According to (\ref{equ:CAF}), the PRSL at the Doppler shift $\theta$ and the Doppler profile at $k=0$  respectively depend on 
\begin{equation}
	\label{equ:F}
	{F_{\mathcal{S,W}}}\left( \theta  \right) = \sum\limits_{m = 0}^{M - 1} {{s_m}{w_m}{{\rm{e}}^{j\theta m}}},\ {G_{\mathcal{W}}}\left( \theta  \right) = \sum\limits_{m = 0}^{M - 1} {{w_m}{{\rm{e}}^{j\theta m}}}.
\end{equation}
By assigning  $F_{\mathcal{S,W}}\left(\theta\right)$ a high-order null point at $\theta=0$, the PRSL can be kept rather low near zero Doppler \cite{Pezeshki2008,Suvorova2007,Dang2011}. The higher the order, the better the Doppler resilience.  The PTM method \cite{Pezeshki2008} achieves a $K$-th order null point by setting $\bf s$ to the PTM code of length $M=2^K,K\in {\mathbb N}^+$, and ${\bf w}={\bf 1}_M$. As a result, it has the highest SNR and Doppler resolution, and however, performs poor in Doppler resilience and Doppler sidelobe suppression. The BD method achieves a $\left(M-1\right)$-th order null point by choosing the standard order and binomial weights, i.e., $w_m=\alpha C_{M-1}^m$. Therefore, it has much better Doppler resilience and lower PDSL, while its degradation in SNR and Doppler resolution could be~acceptable.  Based on the above work,  we develop a novel method to design DRCWs with multiple RSBAs and balanced performance metrics with the aid of numerical optimization theory.

We constrain $F_{\mathcal{S,W}}\left(\theta\right)$ to possess a $K_0$-th order null point at zero Doppler, and a $K_i$-th order null point at $\theta_i,i=1,\cdots,P$. Obviously, it should be satisfied that
\begin{equation}
	\label{equ:K}
	K = {K_0} + 2\sum\limits_{i = 1}^P {{K_i}}  \le M - 1
\end{equation}
Let ${\rm e}^{j\theta}$ and $s_mw_m$ in (\ref{equ:F}) be replaced with $z$ and $y_m$,~respectively. Then, (\ref{equ:F}) can be reformulated as an $(M-1)$-th polynomial with respect to $z$
\begin{equation}
	\label{equ:Fy}
	{F_{\mathcal Y}}\left( z \right) = \sum\limits_{m = 0}^{M - 1} {{y_m}{z^m}} = \sum\limits_{m = 0}^K {{a_m}{z^m}} \sum\limits_{n = 0}^{M - K - 1} {{b_n}{z^n}}
\end{equation}
where ${\bf{a}}=\left[a_0,\cdots,a_K\right]^T$ is given by 
\begin{equation}
	\label{equ:a}
	\sum\limits_{m = 0}^{K } {{a_m}{z^m} = } {\left( {1 - z} \right)^{{K_0}}}\prod\limits_{i = 1}^{{P}} {\left( {1 - z\cos \theta_i  + {z^2}} \right)^{K_i}}, 
\end{equation}
and ${\bf b}\!=\!\left[b_0,\!\cdots,\!b_{M-K-1}\right]^T\!\in\! {\mathbb R}^{M-K}$ is an optimizable~variable. Based on (\ref{equ:Fy}), ${\bf y}=\left[y_0,\cdots,y_{M-1}\right]^T$ can be written as     
\begin{equation}
	\label{equ:y}
	{\bf{y}} = {\bf{a}} \otimes {\bf{b}} = {\bf{Ab}} = { \bf \bar A}{\bf b}
\end{equation}
where $\otimes$ denotes the linear convolution operation, and ${\bf A} \in {\mathbb R}^{M \times (M-K)}$ is a nonsymmetric Toeplitz matrix determined by the first row ${\bf r}^T\!=\!\left[a_0, {\bf 0}_{M-K-1}^T\right]$ and the first column  ${\bf c}\!=\!\left[{\bf a}^T, {\bf 0}_{M-K-1}^T\right]^T$. Orthonormalizing the column vectors of the full-column-rank matrix $\bf A$ to get the semi-orthogonal matrix ${\bf \bar A}$, which satisfies ${{{\bf{\bar A}}}^T}{\bf{\bar A}} = {{\bf{I}}_{M - K}}$. We also impose the energy constraint on $\bf y$, which says
\begin{equation}
	\label{equ:y2}
	{\left\| {\bf{y}} \right\|^{\rm{2}}} = {\left\| {\bf{b}} \right\|^{\rm{2}}} = M.
\end{equation} 

A simple but effective method for the Doppler sidelobe~suppression of coherent pulse trains is to weight the received pulses with a window function. The side-effects of that are reduced SNR and broadened mainlobe. Common window functions includes Hanning, Hamming, Blackman, {\it{etc.}} For a given window function $\bf \bar w$ with ${\left\| {\bf{\bar w}} \right\|^{\rm{2}}}=M$, we use the Euclidean distance to evaluate its similarity to ${\rm{abs}}\left(\bf y\right)$
\begin{equation}
	{\left\| {{\rm{abs}}\left(\bf y\right) - {\bf{\bar w}}} \right\|^2}=2M-{\bf{\bar w}}^T{\rm{abs}}\left(\bf y\right).
\end{equation}
Maximizing the similarity under the constraints (\ref{equ:y}) and (\ref{equ:y2}) results in the following problem
\begin{subequations}
	\label{equ:problem}
	\begin{align}
		{\rm{max}}\quad  & {\bf \bar w}^T{\rm{abs}}\left(\bf \bar Ab\right)\\
		{\rm{s.t.}}\quad & {\left\| {\bf{b}} \right\|^{\rm{2}}} = M.
	\end{align}
\end{subequations}
Notice that the constraint (\ref{equ:y}) is implied in the objective~function. By choosing different windows as $\bf \bar w$, one can emphasize different metrics. Particularly, when $\bf \bar w$  is the rectangular window, problem (\ref{equ:problem}) actually maximizes the SNR.     

Problem (\ref{equ:problem}) is non-convex and discontinuous, and thereby unsolvable in polynomial time. However, a sub-optimal solution of (\ref{equ:problem}) is much easier to find. Introducing the binary auxiliary variable $\bf s$, (\ref{equ:problem}) can be equivalently written as
\begin{subequations}
	\label{equ:problem2}
	\begin{align}
		{\rm{max}}\quad  & {\bf s}^T{\rm{Diag}}\left(\bf \bar w\right) \bf \bar Ab\\
		{\rm{s.t.}}\quad & {\left\| {\bf{b}} \right\|^{\rm{2}}} = M, \ {\bf s} \in \left\{-1,1\right\}^M.
	\end{align}
\end{subequations}
Utilizing the Cauchy-Schwartz inequality, we can calculate the optimal $\bf b$ corresponding to each given $\bf s$
\begin{equation}
	\label{equ:b}
	{\bf{\hat b}} = \alpha {{\bf{\bar A}}^T}{\rm{Diag}}\left(\bf \bar w\right){\bf{s}}, \ \alpha  =  \frac{\sqrt M}{{\left\| {{{\bf{\bar A}}^T}{\rm{Diag}}\left(\bf \bar w\right){\bf{s}}} \right\|^2}}.
\end{equation}
Replacing $\bf b$ in (\ref{equ:problem2}) with expression (\ref{equ:b}), we equivalently transform problem (\ref{equ:problem2}) into the two-way partitioning problem  
\begin{equation}
	\label{equ:problem3}
	\begin{split}
		\max{}& \quad  {{\bf{s}}^T}{\bf{\tilde As}}\\
		{\rm{s.t.}}{}& \quad s_m^2 = 1, \ m = 0, \cdots ,M - 1
	\end{split}
\end{equation}
where ${\bf{\tilde A}}={\rm{Diag}}\left(\bf \bar w\right){\bf{\bar A }}{\bf{\bar A }}^T{\rm{Diag}}\left(\bf \bar w\right)$. With semidefinite programming (SDP) relaxation applied to (\ref{equ:problem3}), we have the following convex problem   
\begin{equation}
	\label{equ:problem4}
	\begin{split}
		\max{}         & \quad  {\rm{tr}}\big( {{\bf{\tilde AS}}}\big)\\
		{\rm{s.t.}}{}  & \quad  {\rm{diag}}\big( {\bf{S}} \big) = {{\bf{1}}_M}\\
		& \quad  {\bf{S}}  \succeq {{\bf{0}}}
	\end{split}
\end{equation}
The solution of (\ref{equ:problem4}), denoted by $\bf \hat S$, can be easily~computed by off-the-shelf optimization toolboxes like CVX. As (\ref{equ:problem4}) is relaxed by omitting the rank-1 constraint on $\bf S$, ${\rm{tr}}\big({{\bf{\tilde A \hat S}}}\big)$ is an upper bound to the optimal value of (\ref{equ:problem3}). A suboptimal solution $\bf \hat s$ of (\ref{equ:problem3}) need to be extracted from $\bf \hat S$. That should be considered in two separate cases: the $\bf \hat S$ whose rank is~1 and the others. The former rarely happens. The latter can be handled with the Gaussian randomization method in \cite{Goemans1995}, which is intended to approximately solve the maximum cut problem with a optimal guarantee of 0.88. See Algorithm~\ref{alg: GR} for the details of generating $\bf \hat s$ from $\bf \hat S$, and Algorithm~\ref{alg: DRCW} for the complete steps of the proposed method. The computational complexity of Algorithm~\ref{alg: DRCW}, which depends on solving SDP, is ${\mathcal O}\left(N^{3.5}{\rm{log}\left(1/\varepsilon\right)}\right)$ at a given SDP solution accuracy $\varepsilon > 0$. 
\begin{figure}[!t]
	\centering
	\subfloat[]{\includegraphics[height=25mm]{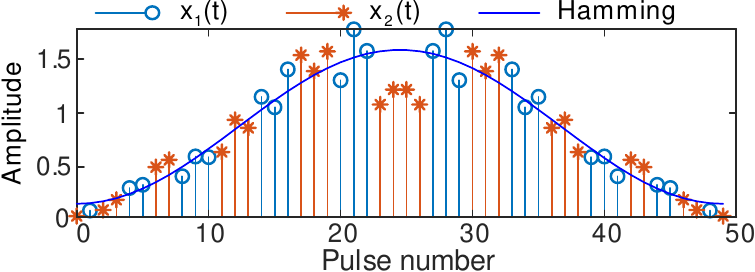}\label{fig:wu1-1}}
	\hfil
	\subfloat[]{\includegraphics[height=25mm]{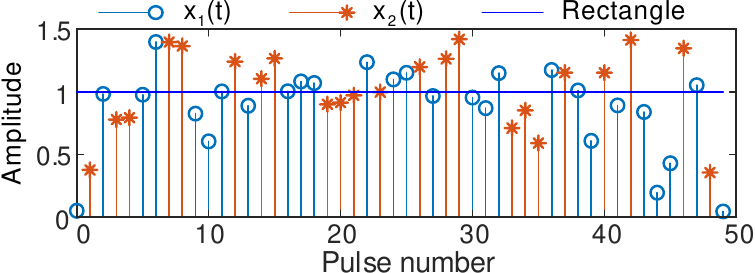}\label{fig:wu1-2}}
	\caption{The pulse trains and pulse weights designed by NM-DRCW with $\bf \bar w$ being the (a)~Hamming window; (b)~rectangular window.}
	\label{fig:wu1}
\end{figure}    
\begin{algorithm}
	\caption{Generating $\bf \hat s$ from $\bf \hat S$}
	\label{alg: GR}
	\begin{algorithmic}[1]
		\STATE Compute the eigenvalues $\lambda_1, \cdots,\lambda_M$ (in descending order) and corresponding eigenvectors ${\bf v}_0,\cdots,{\bf v}_{M-1}$ of $\bf \hat S$;
		\STATE If ${{{\lambda _{\rm{0}}}}} / \sum\nolimits_{m = 1}^{M-1} {{\lambda _m}} \ge \mu $ where $\mu$ is a given large number, ${\bf \hat s}={\rm{sign}}\left({\bf v}_0\right)$, otherwise, turn to step 3;     
		\STATE Generate an random vector $\bf r$ whose entries $r_0,\cdots,r_{M-1}$ drawn from the standard normal distribution; If ${\bf r}^T{\bf \hat s}_M \ge 0$, set $\hat s_m=1$; otherwise, $\hat s_m=-1$.  
	\end{algorithmic}
\end{algorithm}         
\begin{algorithm}
	\caption{Novel method to construct DRCW (NM-DRCW)}
	\label{alg: DRCW}
	\begin{algorithmic}[1]
		\REQUIRE $M$, $\theta_1, \cdots,\theta_P$, $K_0, \cdots,K_P,K$ and $\bf \bar w$.   
		\ENSURE ${\bf \hat s}$ -- the transmit order; ${\bf \hat w}$ -- the receive weights.
		\STATE Construct $\bf a$ according to (\ref{equ:a});  
		\STATE Let ${\bf A }$ be the Toeplitz matrix with ${\bf r}\!=\!\left[a_0, {\bf 0}_{M-K-1}^T\right]^T$ as its first row and ${\bf c}=\left[{\bf a}^T, {\bf 0}_{M-K-1}^T\right]^T$ as its first column;
		\STATE Let ${\bf \bar A }$ be a matrix consisting of an orthonormal basis for the range of ${\bf A }$, and ${\bf{\tilde A}}={\rm{Diag}}\left(\bf \bar w\right){\bf{\bar A }}{\bf{\bar A }}^T{\rm{Diag}}\left(\bf \bar w\right)$; 
		\STATE Solve problem (\ref{equ:problem4}) to get $\bf \hat S$;
		\STATE Generate $\bf \hat s$ with Algorithm~\ref{alg: GR}; repeated the randomization step multiple times to select the $\bf \hat s$ with the largest ${{\bf{\hat s}}^T}{\bf{\tilde A\hat s}}$; 
		\STATE Compute $\bf \hat b$ according to (\ref{equ:b}), and ${\bf \hat y}={\bf \bar A \hat b}$;
		\STATE Compute ${\bf \hat s}={\rm{sign}}\left({\bf \hat y}\right)$ and ${\bf \hat w}={\rm{abs}}\left({\bf \hat y}\right)$.
	\end{algorithmic}
\end{algorithm}

\begin{figure*}[!t] 
	\centering
	\subfloat[]{\includegraphics[height=40mm]{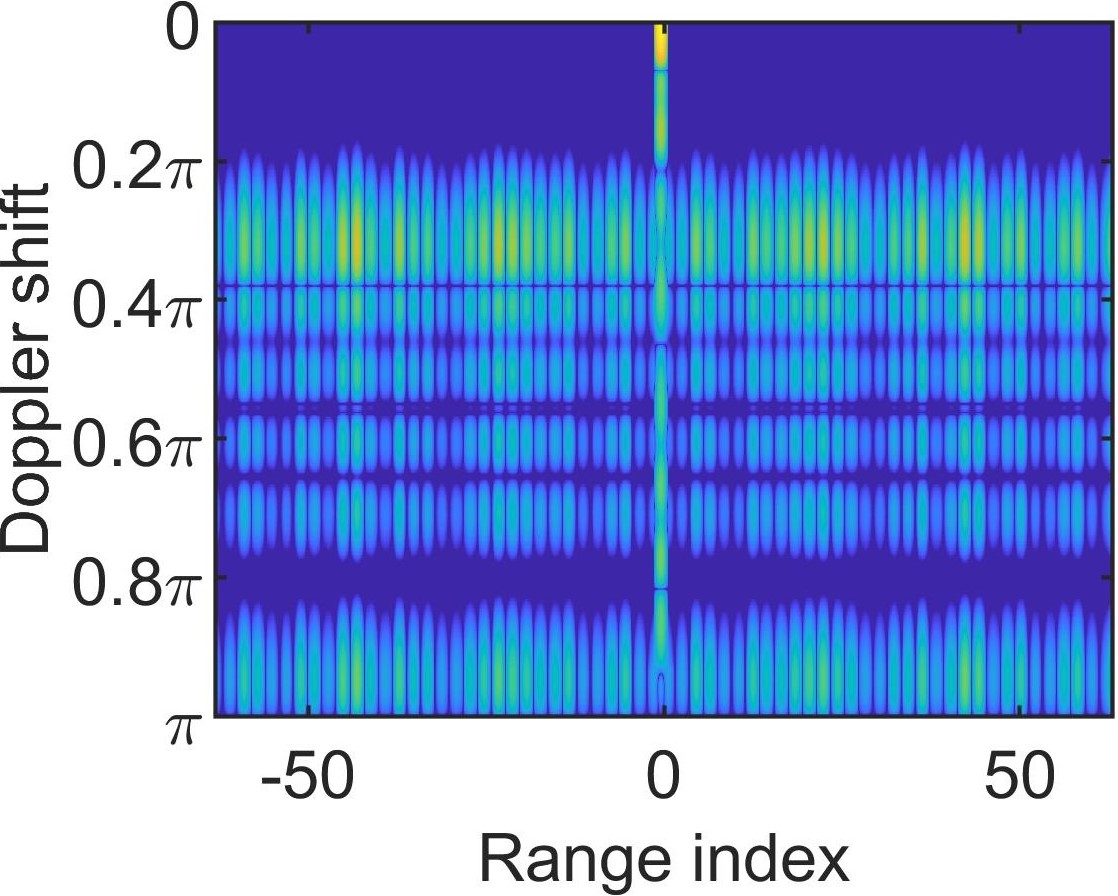}\label{fig:wu2-1}}
	\hfil
	\subfloat[]{\includegraphics[height=40mm]{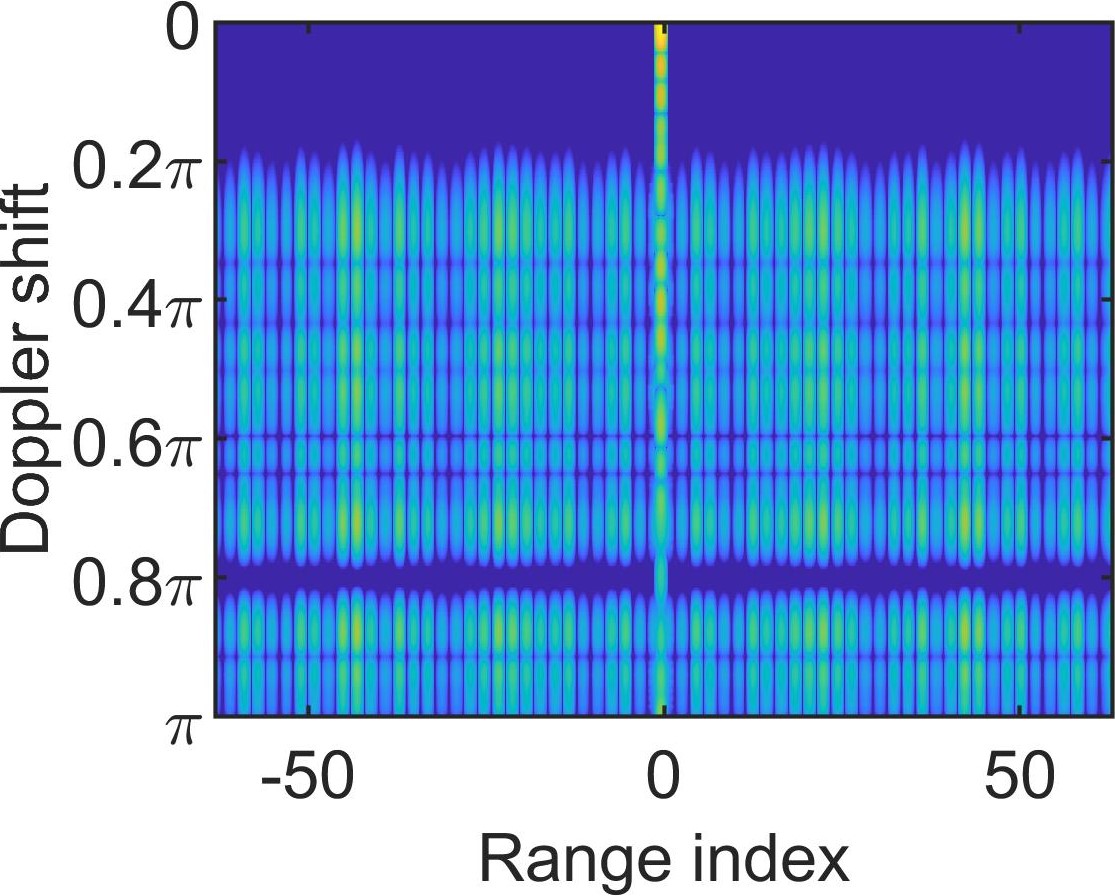}\label{fig:wu2-2}}
	\hfil
	\subfloat[]{\includegraphics[height=40mm]{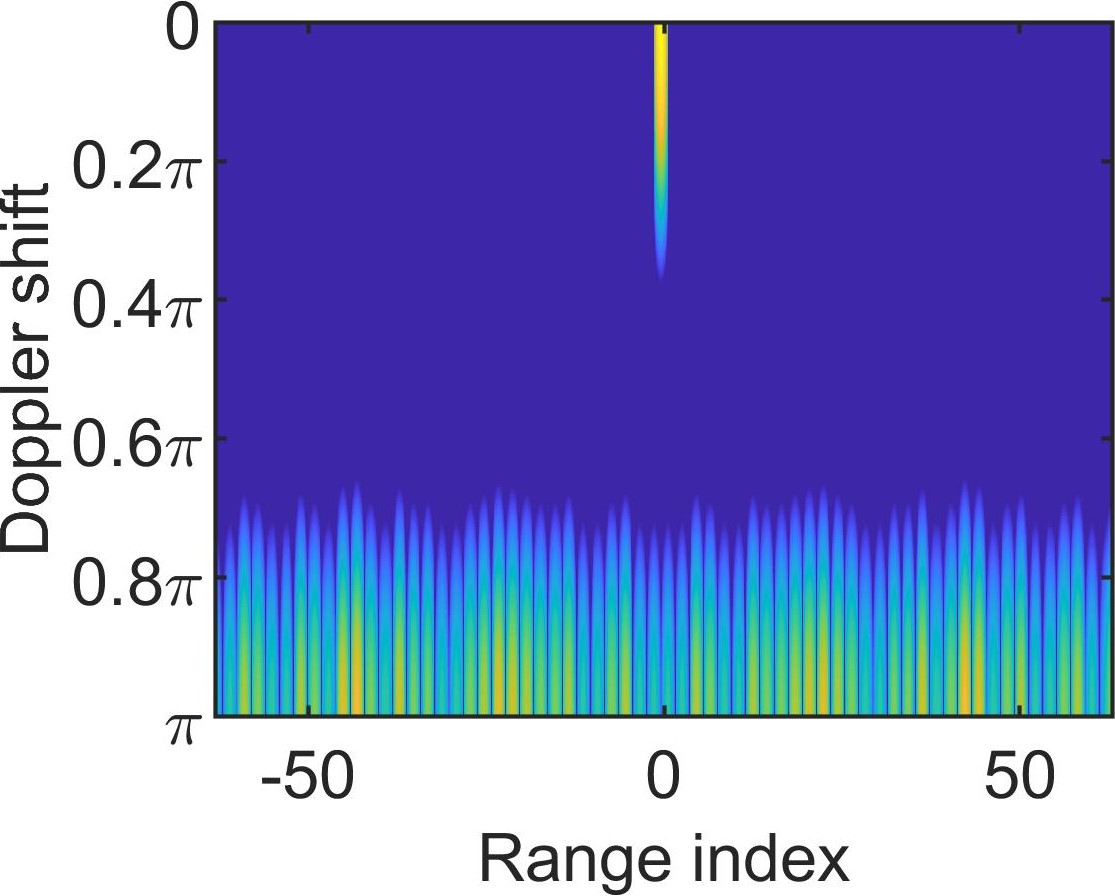}\label{fig:wu2-3}}
	\hfil
	\subfloat{\includegraphics[height=40mm]{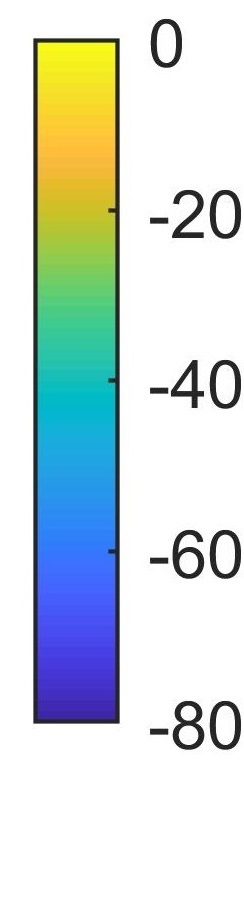}}
	\caption{The normalized cross-ambiguity function of $u_{\mathcal S},u_{\mathcal W}$  designed by (a)~NM-DRCW with $\bf \bar w$=Ham.; (b)~NM-DRCW with $\bf \bar w$=rect.; (c)~the BD method.}
	\label{fig:wu2}
\end{figure*}

\begin{figure*}[!t]
	\centering
	\subfloat[]{\includegraphics[height=40mm]{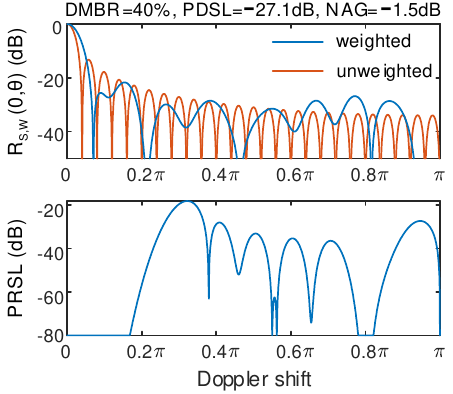}\label{fig:wu3-1}}
	\hfil
	\subfloat[]{\includegraphics[height=40mm]{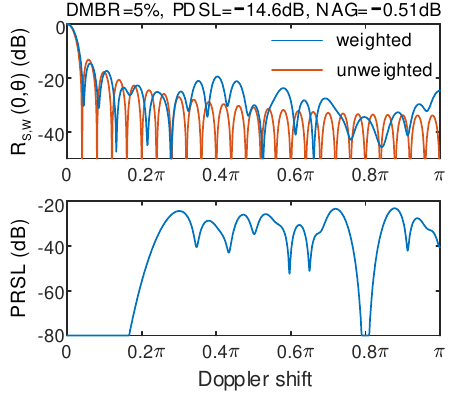}\label{fig:wu3-2}}
	\hfil
	\subfloat[]{\includegraphics[height=40mm]{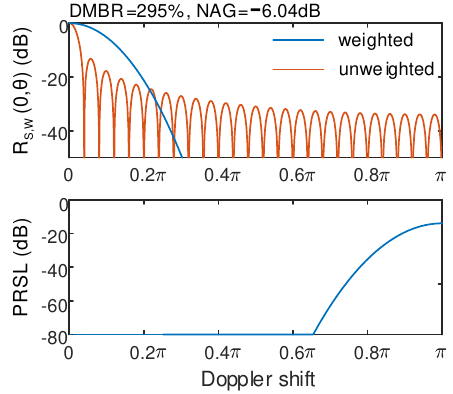}\label{fig:wu3-3}}
	\caption{The Doppler profile (top) and PRSL vs. Doppler shifts (bottom) of the CAF  (a)~$\bf \bar w$=Ham.; (b)~$\bf \bar w$=rect.; (c)~the BD method. (the Doppler profile of the CAF of an unweighted pulse train is also plotted for comparison)}
	\label{fig:wu3}
\end{figure*}

\begin{table*}[t!]
	\centering 
	\begin{threeparttable}
		\renewcommand{\arraystretch}{1.0}   
		\renewcommand{\tabcolsep}{9.0pt}
		\caption{The performance metrics of DRCWs under various Doppler null constraints}              
		\label{tab:sequences}                                     
		\begin{tabular}{ccccccccc}		
			\toprule
			&$K_0$&10&15&20& 25&30&35&40\\
			\midrule
			\multirow{4}{*}{Hamming}&RSBA              &$\left[0,0.08\pi\right]$   &$\left[0,0.14\pi\right]$   &$\left[0,0.20\pi\right]$      &$\left[0,0.26\pi\right]$ 
			&$\left[0,0.35\pi\right]$   &$\left[0,0.42\pi\right]$ &$\left[0,0.50\pi\right]$\\
			&DMBR               &$45\%$                     &$50\%$                     &$45\%$                         &$50\%$  &$55\%$  &$65\%$ &$70\%$\\
			&PDSL               &$-32.6{\,\rm{dB}}$                 &$-30.1{\,\rm{dB}}$               &$-27.1{\,\rm{dB}}$                  &$-20.0{\,\rm{dB}}$ &$-24.2 {\,\rm{dB}}$ &$-18.5{\,\rm{dB}}$ &$-17.8{\,\rm{dB}}$\\
			&NAG               &$-1.50{\,\rm{dB}}$                 &$-1.59{\,\rm{dB}}$               &$-1.56{\,\rm{dB}}$                  &$-1.76{\,\rm{dB}}$ &$-1.91{\,\rm{dB}}$ &$-2.38{\,\rm{dB}}$ &$-2.5{\,\rm{dB}}$\\
			\midrule
			\multirow{4}{*}{Rectangular}&RSBA               &$\left[0,0.07\pi\right]$   &$\left[0,0.13\pi\right]$   &$\left[0,0.19\pi\right]$      &$\left[0,0.25\pi\right]$ 
			&$\left[0,0.32\pi\right]$   &$\left[0,0.40\pi\right]$ &$\left[0,0.48\pi\right]$\\
			&DMBR               &$<1\%$                     &$<1\%$                     &$5\%$                         &$5\%$  &$20\%$  &$25\%$ &$45\%$\\
			&PDSL                 &$-13.8{\,\rm{dB}}$                 &$-14.5{\,\rm{dB}}$               &$-14.3{\,\rm{dB}}$                  &$-13.6{\,\rm{dB}}$ &$-14.3{\,\rm{dB}}$ &$-12.6{\,\rm{dB}}$ &$-13.0{\,\rm{dB}}$\\
			&NAG               &$-0.05{\,\rm{dB}}$                 &$-0.23{\,\rm{dB}}$               &$-0.29{\,\rm{dB}}$                  &$-0.65{\,\rm{dB}}$ &$-1.00{\,\rm{dB}}$ &$-1.63{\,\rm{dB}}$ &$-2.31{\,\rm{dB}}$\\
			\bottomrule
		\end{tabular}
	\end{threeparttable}
\end{table*}                       

\section{Numerical Result}
In our simulations, the radar waveform parameters are set as follows: the~sequence length $N=64$, the pulse number $M=50$ and the duty cycle of the transmitter ($NT_c$ divided by $T$) is $5\%$. The background contains many stationary strong interference like buildings, among which there exist multiple slightly moving targets with Doppler shifts below $0.2\pi\,\rm{rad}$, and high-speed targets whose Doppler shifts are about $0.8\pi\,\rm{rad}$. Setting $\theta_0=0,\theta_1=0.8\pi$ and $K_0=20,K_1=4$,  we construct two DRCWs with the Hamming and the rectangular window as the template respectively. The pulse trains and pulse weights designed by NM-DRCW are plotted in Fig.1. The normalized cross-ambiguity function of $u_{\mathcal S}$ and $u_{\mathcal W}$  designed by our method and the BD method are shown in Fig.~\ref{fig:wu2}. The Doppler profiles at $k=0$ and PRSLs versus Doppler shifts are given in Fig.~\ref{fig:wu3}. We observe that: i)~The BD method fails to provide required Doppler resilience, and suffers significant decrease in SNR and Doppler resolution. ii)~NM-DRCW can satisfy our requirement by setting multiple RSBAs, and the side effects are acceptable; iii)~NM-DRCW with the Hamming window has the Doppler sidelobe inhibited lower than that with the rectangular window, while the latter performs betters in SNR and Doppler resolution than the former. Setting $K_1=0$ and altering $K_0$ from 10 to 40, we construct DRCWs under various conditions and detail the performance metrics in TABLE~I. It shows that flexible compromises among different aspects of DRCWs can be made by NM-DRCW.             

\section{Conclusion}
In this paper, we propose a novel and flexible method for DRCW construction. Our method surpasses the existing ones because it can achieve better Doppler resilience, and meanwhile take other aspects of waveform performance into account. In terms of both effectiveness and practicability, the DRCWs thus obtained are more suitable for detecting weak slightly-moving targets in a dense-interference environment than the waveforms with locally-shaped ACFs. DRCWs can thoroughly prohibit the range sidelobes at all lags, and greatly simplify or omit the procedures of environmental perception, real-time waveform designing, scheduling and transmitting in a cognitive radar.

\end{document}